# Frequency Domain Approach for Activity Classification using Accelerometer

Wan-Young Chung, Amit Purwar and Annapurna Sharma, *Member, IEEE*

*Abstract*— Activity classification was performed using MEMS accelerometer and wireless sensor node for wireless sensor network environment. Three axes MEMS accelerometer measures body's acceleration and transmits measured data with the help of sensor node to base station attached to PC. On the PC, real time accelerometer data is processed for movement classifications. In this paper, Rest, walking and running are the classified activities of the person. Both time and frequency analysis was performed to classify running and walking. The classification of rest and movement is done using Signal magnitude area (SMA). The classification accuracy for rest and movement is 100%. For the classification of walk and Run two parameters i.e. SMA and Median frequency were used. The classification accuracy for walk and running was detected as 81.25% in the experiments performed by the test persons.

## I. INTRODUCTION

Major increment in the elderly population has created the necessity to change the traditional health care system. The activity and behavior monitoring of elderly can give significant information about their health and help in managing weight, in lowering blood pressure, in increasing the level of the good high-density lipoprotein (HDL) cholesterol [1].

*Juha Parkka et al.* [1] focus on the data collected from the various sensors worn on the person's body. The main aim of this work was to study activity classification, which are the most information-rich sensors and what kind of signal processing and classification methods should be used for activity classification. Communication of the sensor unit to the main PC was not a concern of this research. Several Time domain and frequency domain features were selected for activity classification. The most recent work is presented by Roozbeh Jafarie et al. [2] which used the Singular value decomposition (SVD) for structural pattern recognition with angle from the vertical axes. For activity classification neural networks based classifier was used. Radio communication was based on Bluetooth module. Researches like Speedy [3] used simple acceleration magnitude and velocity peaks to detect the fall and other movement activity with moderate accuracy. In this research sensor were worn on wrist which caused extra noise due to hand movement.

Development of sensor networks which consist of small, low-power, and low-cost devices with limited computational and wireless communication capabilities called sensor node or motes [4] can replace big and costly device like PDA and mobile phones for wireless communication in healthcare. These sensor nodes operate on a component-based runtime environment called TinyOS [5] that has been specifically designed to provide support for deeply embedded systems with a minimal amount of physical hardware.

In this research these motes were used to monitor the activity of person in the home or near home environment. The Activities considered for classification are Rest, Walk and Run. Both time and frequency domain parameters were used to classify the activities.

## II. SYSTEM DESIGN

### A. Hardware Design

Telos type sensor node, TIP710 (Maxfor Co. Ltd., Korea) was used as computation and communication resource. Apart from Texas Instruments microcontroller MSP430, TIP 710 has CC2420 2.4 GHz ISM band radio capable of data transfer at 250 kb/s. Capacitive type Microelectromechanical Sensors (MEMS) triaxial accelerometer MMA 7260 (Freescale Inc., USA) was used in this research. Accelerometer sensor has a range of -6g to 6g and sensitivity of 200mV/g, g is here the acceleration due to gravity. The whole sensor unit is powered by two AA batteries.

### B. Software Design

The software for sensor unit was developed using nes-C as a programming language with Tiny-OS [5] as the real time operating system to make it compatible with sensor network. Tiny-OS handles various resources like memory and processing power and keep synchronization between various events. Priority was given to the hardware interrupt like interrupt from ADC or radio over the data processing task. Tiny-OS is a component based event driven operating system in which tasks perform primary work and these tasks are atomic, i.e. task can not be preempted by one another they run up to completion, with respect to each other but can be interrupted by events. Events are the hardware interrupts. Software architecture is based on Active Message

Manuscript received April 7, 2008.
Wan-Young Chung, Dr., is with the Division of Computer & Information Engineering, Dongseo University, Busan 617-716, Korea (phone: +82-51-320-1756; fax: +82-51-327-8955; e-mail: wychung@dongseo.ac.kr).
Amit Purwar, is with Dept. of Ubiquitous IT, Graduate School of Design & IT, Dongseo University, Busan 617-716, Korea (e-mail: purwaramit@gmail.com).
Annapurna Sharma, is with Dept. of Ubiquitous IT, Graduate School of Design & IT, Dongseo University, Busan 617-716, Korea (e-mail: sharmaannapurna@gmail.com).

communication model [6] because it is based on the concepts of combining communication and computation and matching them to the hardware capabilities in concurrent event based operations. At lower level components have handlers connected directly to hardware interrupt which can be external interrupts, timer events, or counter events.

III. METHODS

*A. Calibration & Preprocessing*

At the sensor unit, analog acceleration data which is the continuous voltage signal from triaxial accelerometer sensor, were sampled and quantized for radio transmission. So a calibration was required before any data processing to receive the meaningful and correct data (in terms of gravity acceleration) from V (mV) values. The device is calibrated using linear calibration method by considering device values both at 0g and 1g condition. 'g'- is acceleration due to gravity in $m/s^2$. In this method, data values in mV were collected for device position which result acceleration values of +1g, 0g and -1g for each axis. For the calibration, the prototype sensor unit was worn by the test person at the chest and the 3-axis data was collected for horizontal upright position and horizontal inverted position for an approximate interval of 5 seconds. The calibration is so done as the measured acceleration is '+1g' when the device is positioned upright horizontal and '-1g' when the device is positioned inverted horizontal.

Selection of sampling frequency is an important design consideration for acquiring meaningful data. Sampling frequency should fulfill the Nyquist criteria [7] in order to get a proper signal reconstruction at the base station. Initially, Frequency response of the z-axis data during running is analyzed to find the maximum frequency component of the signal. Most of the signal strength was found to lie within 15 Hz (shown in fig 1 for raw z-axis data) and maximum peaks to discriminate the activities up to a range of 5 Hz as shown in fig. 1. So 50 Hz was taken as the sampling frequency to fulfill nyquist criteria.

The calibrated data, at the base station, has some noise spikes in it. So a Moving average filter, of window size 3, was used to remove the random sharp noise spikes in time domain. This filter smoothened the signal after removing the sharp peaks. Acceleration data received from the three axis accelerometer have mainly two components, one is Body movement component due to linear motion of the person wearing sensor unit and another is gravity component which is always present because of the acceleration due to gravity. For classifying human activities like rest, moving, walking and running, only body movement component is of importance. Since these two components overlap in both time and frequency domain so it is not easy to separate these components, However filtering can be used to partially separate these two components. Normally gravity component signal has lower frequency range in between 0-0.8 Hz. So an Elliptical IIR High Pass filter (HPF) of seventh order with 0.5 Hz cutoff frequency was used to separate the bodily accelerations from the gravity accelerations.

*B. Detection Algorithm*

The magnitude of acceleration differs for different activities like rest, walking and running. Running shows higher value of acceleration magnitude of 3-axis data over walking and rest. But this magnitude acceleration can overlap if person is running slowly. So for the classification between running and walking, both the time and frequency domain features need to be considered. For deriving the frequency domain features Fourier transform and Power spectral density (PSD) are generally used. The PSD gives a variation of signal strength versus frequency while the Fourier transform gives the variation of signal magnitude versus frequency. We used the power spectral density as it provides more elaborate view of the signal versus frequency than the simple Fourier Transform. Calculations of Power Spectrum Density were performed using Welch [8] method. This method is an improved estimator of PSD in terms of spectral leakage over the periodogram method. PSD curves for walking and running activities are shown in Fig. 2. It is showing how the Signal strength for walking and running is distributed differently in frequency domain. The median frequency was calculated from power spectral density (PSD) curve of the Z-axis preprocessed data. It is the frequency which divides the area under the PSD curve in two equal halves. Median frequency ($f_m$) was calculated using Average Power ($P_{av}$) derived from PSD. Area under the PSD curve represents the total average signal power ($P_{av}$) and is given by Eq. (1)

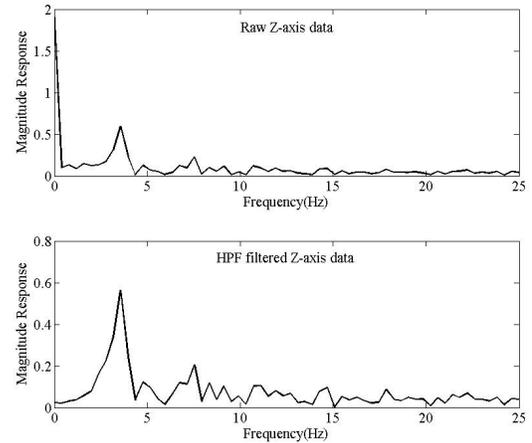

Fig. 1. Magnitude Responses of Raw and Filtered Z-axis data for Running.

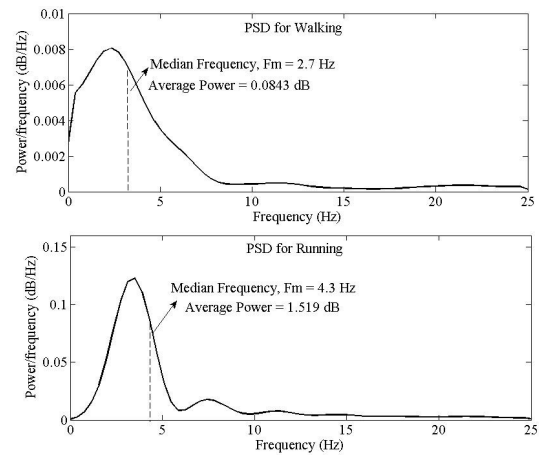

Fig. 2. Power spectral density (PSD) and median frequency for the Z-axis data for walking and running

$$P_{av} = f_s \Sigma P_{xx} / \text{Length of } P_{xx} \qquad (1)$$

Where $f_s$ is the sampling frequency of the signal, $P_{xx}$ is the PSD value in dB/Hz for a particular frequency and Length of ($P_{xx}$) is the number of points for which $P_{xx}$ is calculated. Fig 2 is also showing the values of the median frequencies calculated from PSD curves for the walking and running. As the median frequency differs for the two activities, it is used in the algorithm for classifying walking and running.

For distinguishing between the rest and movement (either walk or run) the time domain feature Signal magnitude area

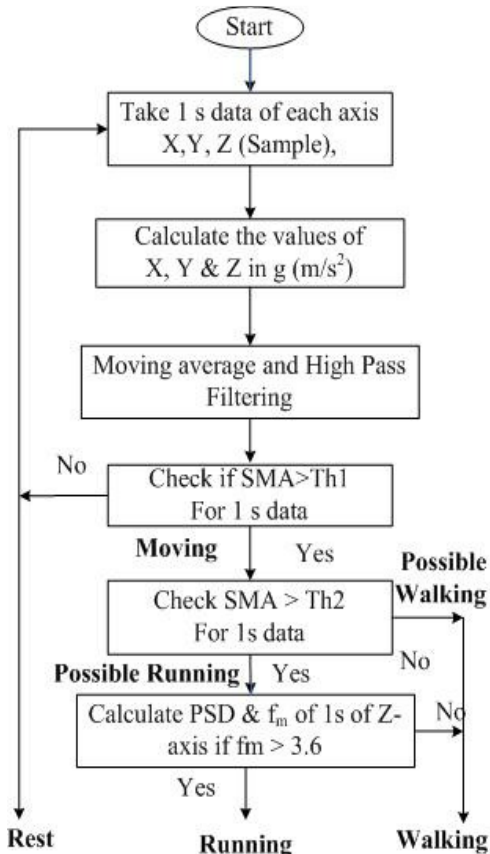

Fig. 3. Flow chart of the algorithm for activity classification.

(SMA) was derived. Signal magnitude area was calculated using the area under the magnitude of the RMS of all the three axis preprocessed acceleration data of 1 second duration. This SMA is calculated after dividing area under the acceleration curve with the time interval. In the detection algorithm, firstly motion and rest classification is performed with the help of first threshold Th1 for signal magnitude area (SMA) for 1 second data. If SMA for 1 second data crosses the fixed threshold then further another threshold Th2 is checked for walking and running. Generally SMA for running is higher than walking, but in some cases this SMA value overlaps so median frequency was used to finally conclude the classification. Fig. 3 shows the steps involved in the algorithm for detection.

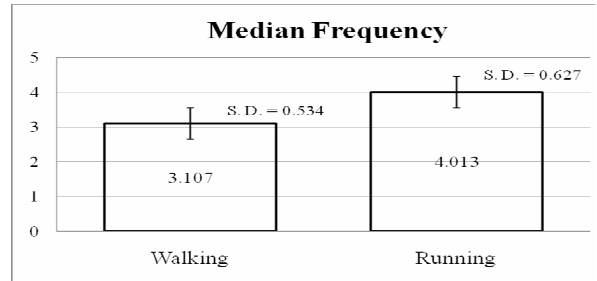

Fig. 4. Bar Graph showing Median Frequency and its Standard Deviation.

IV. EXPERIMENTS & RESULTS

Experiments were performed on test person by wearing the prototype sensor unit on the chest as per the setting stated in calibration and the preprocessing part of the paper. The sampling frequency at the sensor node was taken as 50Hz. Total 10 experiments were performed in which the test person was asked to perform Rest (Stand), walk and run activities for 5 minutes each, in the order stated only. Test person was accompanied by an annotator, who marked the activities and interval for reference purposes. This marking of activity

TABLE 1.
PERFORMANCE RESULTS OF WALKING & RUNNING CLASSIFICATION

| Activity | $N_t$ | $D_t$ | $f_m$Min (Hz) | $f_m$Max (Hz) |
|---|---|---|---|---|
| Walk | 24 | 20 | 2.3 | 3.51 |
| Run | 24 | 19 | 3.12 | 5 |
| Walk+Run | 48 | 39 | - | - |

$N_t$ = Total Number of Experiments, $D_t$ = Number of correctly detected activities, $f_m$Min = Minimum detected median frequency, $f_m$Max = Maximum median frequency during experiments, is used for verification of activity classification by algorithm. Real time activity data of the entire three axes during each experiment were recorded for signal processing and activity classification. Table 1 shows the performance results of the algorithm for walking and running classification. Classification results show an accuracy of 81.25 %. Bar graph of the median frequency is plotted to view the standard deviation in the $f_m$ in the different experiments as shown in fig. 4.

## V. Conclusion

Activity classification system was implemented with triaxial accelerometer and wireless sensor node to be used in wireless ad-hoc communication environment. Accuracy of the overall classification was detected as 81.25%. Detection accuracy shows the feasibility of the system for practical implementation. System is compatible for ubiquitous health using sensor network due to use of sensor node. The feature parameters used in the detection algorithm for activity classification includes signal magnitude area of three axes acceleration data and the median frequency derived from power spectrum density of the Z-axis signal. For analyzing both the time and frequency domain parameters simultaneously, wavelet transform can also be used further.

## References

[1] Juha Parkka, Miikka Ermes, Panu Korpipaa, Jani Mantyjarvi, Johannes Peltola and Ilkka Korhonen, "Activity Classification Using Realistic Data From Wearable Sensors", *IEEE Transactions on Information Technology in Biomedicine,* vol.10 no.1, 2006, pp 119-128.

[2] Roozbeh Jafari, Wwnchao Li, Ruzena Bajcsy, Steven Glaser, and Shankar Sastry, "Physical Activity Monitoring for Assisted living at home", *4th International Workshop on Wearable and Implantable Body Sensor Network,* vol. 13, 2007, pp 99-104.

[3] T. Degen, H. Jaeckel, M. Rufer, and S. Wyss, "SPEEDY: a fall detector in a wrist watch", *Proc. Seventh IEEE International Symposium on Wearable Computing,* pp. 184-187, 2003.

[4] Polastre Joseph, Szewczyk Robert, Culler David, "Telos: enabling ultra-low power wireless research", *Fourth International Symposium on Information processing in sensor network,* 2005

[5] J. Hill, R. Szewczyk, A. Woo, S. Hollar, D. Culler, and K. Pister, "System architecture directions for network sensors", *International Conference on Architectural Support for Programming languages and Operating Systems, Cambridge,* November 2000

[6] Philip Buonadonna, Jason Hill, David Culler, "Active Message Communication for Tiny Networked Sensors", *Proceeding of 20th Annual Joint Conference of IEEE*, 2001.

[7] Alan V. Oppenheim, R.W. Schafer, and John R. Buck, Discrete-Time Signal Processing, 2nd Ed., Prentice Hall, p.140, 1998.

[8] Welch, P.D. "The Use of Fast Fourier Transform for the Estimation of Power Spectra: A Method Based on Time Averaging Over Short, Modified Periodograms", *IEEE Trans. Audio Electroacoust.* vol. AU-15. pp. 70-73, June 1967.